\documentclass[twocolumn]{revtex4}

\usepackage{graphicx}

\def\giorno{{Revised version --} 30/06/2020}

\def\a{\alpha}
\def\b{\beta}
\def\ga{\gamma}

\def\s{\sigma}
\def\la{\lambda}

\def\I{\mathcal{I}}

\def\^#1{\widehat{#1}}
\def\wt#1{\widetilde{#1}}

\def\beql#1{\begin{equation} \label{#1}}
\def\beq{\begin{equation}}
\def\eeq{\end{equation}}

\def\<{\langle}
\def\>{\rangle}

\def\({\left(}
\def\){\right)}

\def\[{\left[}
\def\]{\right]}

\def\eqref#1{(\ref{#1})}

\def\EOR{\hfill $\odot$}

\def\REM#1#2{{\small \medskip\noindent{\bf Remark {#1}.} {#2} \EOR } }

\begin{document}

\title{Social distancing versus early detection and contacts tracing in epidemic management}

\author{Giuseppe Gaeta\\
{\it Dipartimento di Matematica, Universit\`a degli Studi di Milano} \\
{\it via Saldini 50, 20133 Milano (Italy)} \\ {\rm and} \\ {\it SMRI,  00058 Santa Marinella (Italy)} \\ {\tt giuseppe.gaeta@unimi.it} }

\date{\giorno}

\begin{abstract}
\noindent
Different countries -- and sometimes different regions within the same countries -- have adopted different strategies in trying to contain the ongoing COVID-19 epidemic; these mix in variable parts social confinement, early detection and contact tracing. In this paper we discuss the different effects of these ingredients on the epidemic dynamics; the discussion is conducted with the help of two simple models, i.e. the classical SIR model and the recently introduced variant A-SIR (arXiv:2003.08720) which takes into account the presence of a large set of asymptomatic infectives.
\end{abstract}

\maketitle

\section{Introduction}

Different countries are tackling the ongoing COVID-19 epidemics with different strategies. Awaiting for a vaccine to be available, the three tools at our disposal are \emph{contact tracing}, \emph{early detection} and \emph{social distancing}. These are not mutually exclusive, and in fact they are used together, but the accent may be more on one or the other.

Within the framework of classical SIR \cite{KMK,Murray,Britton,Heth,Edel} and SIR-type models, one could say (see below for details) that these strategies aim at changing one or the other of the basic parameters in the model.

In this note we want to study -- within this class of models -- what are the consequences of acting in these different ways. We are interested not only in the peak of the epidemics, but also in its duration.

In fact, it is everybody's experience in these days that social distancing -- with its consequence of stopping all kind of economic activities -- has a deep impact on our life, and in the long run is producing impoverishment and thus a decline in living conditions of a large part of population.

In the present study we will not specially focus on COVID, but discuss the matter in general terms and by means of general-purpose models.

Our Examples and numerical computations will however use data and parameters applying to (the early phase of) the current COVID epidemic in Northern Italy, in order to have realistic examples and figures; we will thus use data and parameters arising from our analysis of epidemiological data in the early phase of this epidemic \cite{Gcov,Gasir}\footnote{{In this regard, also in view of the rapid evolution of the pandemic, it should be pointed out that the first version of this paper was submitted at mid-May, and data thus run up to that time. The Appendix contains more recent data.}}. Unavoidably, we will also here and there refer to the COVID case.

Some observations deviating from the main line of discussion {-- or which we want to pinpoint for easier reference to them -- } will be presented in the form of Remarks. The symbol $\odot$ marks the end of Remarks.

\section{The SIR model}

In the SIR model \cite{KMK,Murray,Britton, Heth,Edel}, a population of constant size (this means the analysis is valid over a relatively short time-span, or we should consider new births and also deaths not due to the epidemic) is subdivided in three classes: Susceptibles, Infected (and by this also Infectives), and Removed. The infected are supposed to be immediately infective (if this is not the case, one considers so called SEIR model to take into account the delay), and removed may be recovered, or dead, or isolated from contact with susceptibles.

\REM{1}{We stress that while in usual textbook discussions of the SIR model \cite{Murray,Britton, Heth,Edel} the Removed are either recovered or dead, in the framework of COVID modeling the infectives are \emph{removed} from the infective dynamics  -- i.e. do not contribute any more to the quadratic term in the equations \eqref{eq:SIR} below -- through isolation. This means in practice hospitalization in cases where the symptoms are heavy and a serious health problem develops, and isolation at home (or in other places, e.g. in some countries or region specific hotels were used to this aim) in cases where it is estimated that there is no relevant risk for the health of the infective. In this sense, the reader should pay attention to the meaning of $R$ in the present context.}
\bigskip

The nonlinear equations governing the SIR dynamics are written as
\begin{eqnarray}
dS/dt &=& - \, \a \, S \, I \nonumber \\
dI/dt &=& \a \, S \, I \ - \ \b \, I \label{eq:SIR} \\
dR/dt &=& \b \, I \ . \nonumber \end{eqnarray}
These should be considered, in physicists' language, as \emph{mean field}  equations; they hold under the (surely not realistic) assumption that all individuals are equivalent, and that the numbers are sufficiently large to disregard fluctuations around mean quantities.

Note also that the last equation amounts to a simple integration,
$ R(t) = R_0 + \b \int_{t_0}^t I(y) d y$;
thus we will mostly look at the first two equations in \eqref{eq:SIR}.

We also stress, however, that epidemiological data can only collect time series for $R(t)$: so this is the quantity to be compared to experimental data \cite{Murray}.

\REM{2}{In fact, as stressed in Remark 1, in the case of a potentially dangerous illness (as COVID), once the individuals are identified as infective, they are effectively removed from the epidemic dynamic through hospitalization or isolation.}
\bigskip

According to our equations \eqref{eq:SIR}, $S(t)$ is always decreasing until there are infectives.
The second equation in \eqref{eq:SIR} immediately shows that the number of infectives grows if $S$ is above the \emph{epidemic threshold}
\beql{eq:ga} \ga \:= \ \b / \a \ . \eeq
Thus to stop an epidemic once the numbers are too large to isolate all the infectives, we have three (non mutually exclusive) choices within the SIR framework:
\begin{itemize}
\item[(a)] Do nothing, i.e. wait until $S(t)$ falls below the epidemic threshold;
\item[(b)] Raise the epidemic threshold above the present value of $S(t)$ by decreasing $\a$;
\item[(c)] Raise the epidemic threshold above the present value of $S(t)$ by increasing $\b$.
\end{itemize}

In practice, any State will try to both raise $\b$ and lower $\a$, and if this is not sufficient await that $S$ falls below the attained value of $\ga$.

In order to understand how this is implemented, it is necessary to understand what $\a$ and $\b$ represent in concrete situations.

The parameter $\b$ represents the \emph{removal rate} of infectives; its inverse $\b^{-1}$ is the average time the infectives spend being able to spread the contagion. Raising $\b$ means lowering the time from infection to isolation, hence from infection to detection of the infected state.

The parameter $\a$ represents the \emph{infection rate}, and as such it includes many thing. It depends both on the infection vector characteristics (how easily it spreads around, and how easily it infects a healthy individual who gets in contact with it), but is also depends on the occasions of contacts between individuals. So, roughly speaking, it is proportional to the number of close enough contacts an individual has with other ones per unit of time. It follows that -- if properly implemented -- social distancing results in reducing $\a$.

Each of these two actions presents some problem. There is usually some time for the appearance of symptoms once an individual is infected, and the first symptoms can be quite weak. So early detection is possible only by fast tracing and laboratory checking of all the contacts of those who are known to be infected. This has a moderate cost (especially if compared to the cost of an Intensive Care hospital stay) but requires an extensive organization.

On the other hand, social distancing is cheap in immediate terms, but produces a notable strain of the societal life, and in practice -- as many of the contacts are actually work related -- requires to stop as many production and economic activities as possible, i.e. has a formidable cost in the medium and long run. Moreover, it cannot be pushed too far, as a number of activities and services (e.g. those carrying food to people, urgent medical care, etc.) can not be stopped.

Let us come back to \eqref{eq:SIR}; using the first two equations, we can study $I$ in terms of $S$, and find out that
\beq I \ = \ I_0 \ + \ (S_0 - S) \ - \ \ga \ \log (S_0 / S) \ . \eeq
As we know that the maximum $I_*$ of $I$ will be reached when $S=\ga$, this allows immediately to determine the \emph{epidemic peak}. In practice, $I_0$ is negligible and for a new virus $S_0$ corresponds to the whole population, $S_0 =N$; thus
\beql{eq:I*} I_* \ = \ N \ - \ \ga \ - \ \ga \ \log (N/\ga) \ . \eeq
Note that only $\ga$ appears in this expression; that is, raising $\b$ or lowering $\a$ produces the same effect as long as we reach the same $\ga$.

On the other hand, this simple formula does not tell us \emph{when} the epidemic peak is reached, but only that it is reached when $S$ has the value $\ga$. But if measures are taken, these should be effective for the whole duration of the epidemic, and it is not irrelevant -- in particular if the social and economic life of a nation is stopped -- to be able to evaluate how long this will be for.

\subsection{Timescale of SIR dynamics}

\begin{figure}
\centering
  % Requires \usepackage{graphicx}
  \includegraphics[width=200pt]{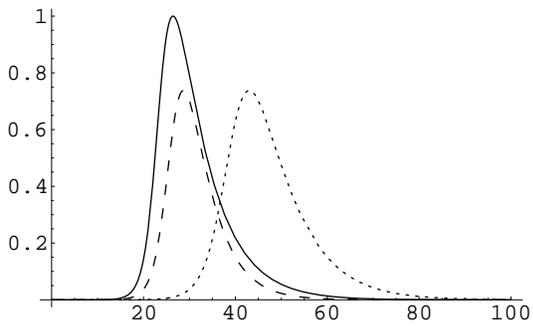}\\
  \caption{Different effect of acting on the $\a$ or the $\b$ parameter. The SIR equations \eqref{eq:SIR} are numerically integrated and $I(t)$ plotted in arbitrary units for given initial conditions and $\a,\b$ parameters (solid), the maximum $I_*$ being reached at $t=t_*$. Then they are integrated for the same initial condition but raising $\b$ by a factor $\vartheta = 3/2$ (dashed) with maximum $I_\b = r I_*$ reached at time $t_\b = \s_\b t_*$; and lowering $\a$ by the same factor $\vartheta = 3/2$ (dotted) with maximum $I_\a= \I_\b$ reached at time $t_\a = \s_\a t_*$. Time unit is one day, $\a = (4/3)*10^{-8}$, $\b = 1/7$; these parameters arise from our fitting of data from the early phase of COVID epidemics in Northern Italy \cite{Gasir}; the population of the most affected area in the initial phase is about 20 million, that of the whole Italy is about 60 million. The numerical simulation is ran with $N = 6*10^7$; it results $r = 0.74$, $\s_\a = 1.63$, $\s_\b = 1.09$, $I_* = 3.08 * 10^7$, $t_* = 26.4$; note that $\s_\a/\s_\b = 3/2 = \vartheta$.}\label{fig:time}
\end{figure}

Acting on $\a$ or on $\b$ to get the same $\ga$ will produce different timescales for the dynamics; see Figure \ref{fig:time}, in which we have used values of the parameters resulting from our fit of early data for the Northern Italy COVID-19  epidemic \cite{Gasir}.

This observation can be made more precise considering the scaling properties of \eqref{eq:SIR}. In fact, consider the scaling
\beql{eq:scale} \a \to \la \,\a \ , \ \ \b \to \la \, \b \ , \ \ t \to \la^{-1} \, t \ . \eeq
It is clear that under this scaling $\ga$ remains unchanged, and also the equations are not affected; thus the dynamics is the same \emph{but with a different time-scale}.

The same property can be looked at in a slightly different way. First of all, we note that one can write $\a = \b/\ga$; moreover, $\a$ appears in \eqref{eq:SIR} only in connection with $S$, and it is more convenient to introduce the variable \beql{eq:Theta} \vartheta \ := \ S/\ga \ . \eeq
Now, let us consider two SIR systems with the same initial data but different sets of parameters, and let us for ease of notation just consider the first two equations of each. Thus we have the two systems
\begin{eqnarray}
\ga^{-1} \ d \vartheta /dt &=& - \b \, \vartheta \, I \ , \ \
dI/dt \ = \ \b \, (\vartheta - 1) \, I \ ; \label{eq:syst1} \\
\wt{\ga}^{-1} \ d \wt{\vartheta}/dt &=& - \wt{\b} \, \wt{\vartheta} \, \wt{I} \ , \ \
d \wt{I}/dt \ = \ \wt{\b} \, (\wt{\vartheta} - 1 ) \, \wt{I} \ . \label{eq:syst2} \end{eqnarray}

We can consider the change of variables ($\la >  0$)
\beq \wt{\b} = \la \, \wt{\b} := \^\b \ , \ \ t \to \la^{-1} \, t := \tau \ . \eeq With this, \eqref{eq:syst2} becomes
$$ \la \, \wt{\ga}^{-1} \, (d \wt{\vartheta}/d \tau) \ = \ - \la \^\b \, \wt{\vartheta} \, \wt{I} \ , \ \
\la (d \wt{I}/d \tau) \ = \ \la \^\b \, (\wt{\vartheta} - 1 ) \, \wt{I} \ . $$  We can thus eliminate the factor $\la$ in both equations. However, if we had chosen $\la = \wt{\b}/\b$, we get $\^\b = \b$; if moreover $\wt{\ga}= \ga$, the resulting equation is just
\beql{eq:syst2b} \ga^{-1} \, d \wt{\vartheta}/d \tau \ = \ - \b \, \wt{\vartheta} \, \wt{I} \ , \ \
d \wt{I}/d \tau \ = \ \b \, (\wt{\vartheta} - 1 ) \, \wt{I} \ . \eeq
But we had supposed the initial data for $\{S,I\}$ and for $\{ \wt{S}, \wt{I} \}$ (and hence also for $\vartheta$ and $\wt{\vartheta}$) to be the same. We can thus directly compare \eqref{eq:syst2b} with \eqref{eq:syst1}.

We observe that $\{\wt{\vartheta},\wt{I}\}$ have thus exactly the same dynamics  in terms of the rescaled time $\tau$ as $\{\vartheta,I\}$ in terms of the original time $t$. In particular, if the maximum of $I$ is reached at time $t_*$, the maximum of $\wt{I}$ is reached at $\tau_* = t_*$, and hence at
\beq \wt{t}_* \ = \ \la \, \tau_* \ = \ \la \, t_* \ . \eeq

Analytical results on the timescale change induced by a rescaling of the $\a$ and $\b$ parameters have recently been obtained by M. Cadoni \cite{Cadoni}; {see also \cite{BaW}}.

\REM{3}{We have supposed infected individuals to be immediately infective. If this is not the case an ``Exposed'' class should be introduced. This is not qualitatively changing the outcome of our discussion, so we prefer to keep to the simplest setting. (Moreover, for COVID it is known that individuals become infective well before developing symptoms, so that our approximation is quite reasonable.)}

\section{A-SIR model}

One of the striking aspects of the ongoing COVID-19 epidemic is the presence of a large fraction of \emph{asymptomatic infectives} \cite{asy1,asy3,asy4,asy5,asy6,asy7,asy8,asy9,asy10,asy11,asy12,asy13}; note that here we will always use ``asymptomatic'' as a shorthand for ``asymptomatic or paucisymptomatic'', as also people with very light symptoms will most likely escape to clinical detection of COVID -- and actually most frequently will not even think of consulting a physician.\footnote{{It is maybe appropriate to stress that asymptomatic infectiveness should not be confused with pre-symptomatic one. See the discussion below, in Remark 5. }}

\subsection{The model and its parameters}

In order to take this aspect into account, we have recently formulated a variant of the SIR model \cite{Gasir} in which together with known infectives $I(t)$, and hence known removed $R(t)$, there are unregistered infectives $J(t)$ and unregistered removed $U(t)$. Note that in this case removal amounts to healing; so while the removal time $\b^{-1}$ for known infected corresponds to the time from infection to isolation, thus in general slightly over the incubation time $T_i$ (this is $T_i \simeq 5.1 \, \mathrm{days}$ for COVID), the removal time $\eta^{-1}$ for unrecognized infects will correspond to incubation time plus healing time.

In the model, it is supposed that symptomatic and asymptomatic infectives are infective in the same way. This is not fully realistic, as one may expect that somebody having the first symptoms will however be more retired, or at east other people will be more careful in contacts; but this assumption simplifies the analysis,and is not completely unreasonable considering that for most of the infection-to-isolation time $\b^{-1}$ the symptoms do not show up.

The equations for the A-SIR model \cite{Gasir} are
\begin{eqnarray}
dS/dt &=& - \, \a \ S \, (I +J) \nonumber \\
dI/dt &=& \a \, \xi \ S \, (I+J) \ - \ \b \, I \nonumber \\
dJ/dt &=& \a \, (1-\xi) \ S \,(I+J) \ - \ \eta \, J \label{eq:ASIR} \\
dR/dt &=& \b \, I \nonumber \\
dU/dt &=& \eta \, J \ . \nonumber \end{eqnarray}
Note that here too we have a ``master'' system of three equations (the first three) while the last two equations amount to direct integrations,
$R(t) = R_0 + \b \int_{t_0}^t I(y) d y$, $U(t) = U_0  + \eta \int_{t_0}^t J(y) d y. $

%\begin{eqnarray*}
%R(t) &=& R_0 \ + \ \b \ \int_{t_0}^t I(y) \ d y \ , \\
%U(t) &=& U_0 \ + \ \eta \ \int_{t_0}^t J(y) \ d y \ . \end{eqnarray*}

The parameter $\xi \in [0,1]$ represents the probability that an infected individual is detected as such, i.e. falls in the class $I$. In the absence of epidemiological investigations to trace the contacts of known infectives, this corresponds to the probability of developing significant symptoms.

In the first (arXiv) circulated version {\cite{Gasir0} of} our previous work \cite{Gasir0}, some confusion about the identification of the class $J$ was present, as this was sometimes considered to be the class of asymptomatic infectives, and sometimes that of not registered ones\footnote{{This confusion was of course corrected in the successive versions of the work, including the published one \cite{Gasir}.}}. While this is not too much of a problem considering the ``natural'' situation, it becomes so when we think of action on this situation.

Actually, and unfortunately, this confusion has a consequence exactly on one of the points we want to discuss here, i.e. the effect of a campaign of chasing the infectives, e.g. among patients with light symptoms or within social contacts of known infectives; let us thus discuss briefly this point.

If $J$ is considered to be the set of asymptomatic virus carriers, then a rise in the fraction of these who are known to be infective, and thus isolated, means that the average time for which asymptomatic infectives are not isolated is decreasing. In other words, we are lowering $\eta^{-1}$ and thus raising $\eta$. On the other hand, in this description $\xi$ is the probability that a new infective is asymptomatic, and this depends only on the nature of the virus and its interactions with the immune system of the infected people; thus in this interpretation $\xi$ should be considered as a constant of nature, and it cannot be changed. (This is the point of view taken in \cite{Gasir}; however some of the assumptions made in {its first version \cite{Gasir0}} were very reasonable only within the concurrent interpretation, described in a moment.)

On the other hand, if $J$ is the class of unknown infectives, things are slightly different. In fact, to be in this class it is needed $(a)$ that the individual has no or very light symptoms; but also $(b)$ that he/she is not traced and analyzed by some epidemiological campaign, e.g. due to contacts with known infected or because belonging to some special risk category (e.g. hospital workers). In this description, $\eta$ is a constant of nature, depending on the nature of the virus and on the response of the ``average'' immune system of (asymptomatic) infected people, while efforts to trace asymptomatic infectives will act on raising the probability $\xi$.

We want to discuss the effect of early detection of infectives, or tracing their contacts, within the second mentioned framework. Note that a campaign of tracing contacts of infectives is useful not only to uncover infectives with no symptoms, but if accompanied by effective isolation of contacts with known infectives, and thus of those who are most likely to be infective, it will also reduce the removal time of ``standard'' (i.e. symptomatic) infectives, possibly to a time smaller than the incubation time itself.

\REM{4}{In this sense, we will look at an increase in $\xi$ as \emph{early detection of infectives}, and at an increase in both $\b$ and $\eta$ (thus a reduction in the removal times $\b^{-1}$ and $\eta^{-1}$) as \emph{tracing contacts of infectives}. This should be kept in mind in our final discussion about the effect of different strategies.}

\REM{5}{As mentioned above, one should also avoid any confusion between asymptomatic and pre-symptomatic infection. In our description, pre-symptomatic infectives -- i.e. individuals which are infective and which do not yet display symptoms, but which will at a later stage display them -- are counted in the class of ``standard'' infectives, i.e. those who will eventually display symptoms and hence be intercepted by the Health system with no need for specific test or contact racing campaigns, exactly due to the appearance of symptoms. Actually one expects that except for the early phase of the epidemics in the countries which were first hit in a given area (such as China for Asia, or Italy for Europe), when symptoms could be attributed to a different illness, most infections by symptomatic people are actually \emph{pre-symptomatic}, as with the appearance of symptoms people are either hospitalized or isolated at home; and even before any contact with the Health system they will avoid contacts with other -- and other people will surely do their best to avoid contacts with anybody displaying even light COVID symptoms. In the case of asymptomatic infectives, instead, unless they are detected by means of a test or contact tracing campaign -- see the forthcoming discussion -- they remain infective until they recover, so that in this case removal is indeed equivalent to (spontaneous) recovery.}

\subsection{A glimpse at COVID matters}

This approach, indeed, was taken in one of the areas of early explosion of the contagion in Northern Italy, i.e. in V\`o Euganeo; this had the advantage of being a small community (about 3,000 residents), and all of them have been tested twice while embargo was in operation. In fact, this was the first systematic study showing that the number of asymptomatic carriers was very high, quite above the expectations \cite{Cri}. Apart from its scientific interest, the approach proved very effective in practical terms, as new infectives were quickly traced and in that specific area the contagion was stopped in a short time.

While testing everybody is not feasible in larger communities, the ``follow the contacts'' approach could be used on a larger scale, especially with the appearance of new very quick kits for ascertaining positivity to COVID.

The model will thus react to a raising of $\xi$ by raising the fraction of $I$ within the class of infectives, i.e. in $K= I+J$; but at the same time, as critical patients are always the same, i.e. represents always the same fraction of $K$, we should pay attention to the fact they will now represent a lower fraction of $I$.
The Chinese experience shows that critical patients are about 10 \% of hospitalized patients (i.e. of those with symptoms serious enough to require hospitalization); and hospitalized patients represented about half of known infected, the other being cured and isolated at home. Similar percentages were observed in the early phase of the COVID epidemic in Italy; the fraction of infectives isolated at home has afterwards diminished, but it is believed that this was due to a different policy for lab exams, i.e. checking prioritarily patients with multiple symptoms suggesting the presence of COVID rather than following the contacts. Actually this policy was followed in most of Italy, but in one region (Veneto) the tracking of contacts and lab exams for them was pursued, and in there the percentages were much more similar to those known to hold for China.

\subsection{Numerical simulations protocol. Parameters and initial data}

In our previous work \cite{Gasir} we have considered data for the early phase of COVID epidemics in Italy, and found that $\b^{-1} \simeq 7$ best fits them while the estimate $\eta^{-1} \simeq 21$ was considered as a working hypothesis. This same work found as value of the contact rate in the initial phase $\a \simeq 1.13*10^{-8}$, and we will use this in our numerical simulations.\footnote{The value of the contact rate was then modified by the restrictive measures adopted by the Government, see the discussion in \cite{Gasir}.}

It should be stressed that the extraction of the parameter $\a$ from epidemiological data is based on the number $S_0 \simeq N$ of susceptibles at the beginning of the epidemic, thus $\a$ and hence $\ga$ depend on the total population. The value given above was obtained considering $N=2*10^7$, i.e. the overall population of the three regions (Lombardia, Veneto and Emilia-Romagna) which were mostly affected in the initial phase.

Our forthcoming discussion, however, does \emph{not} want to provide a forecast on the development of the COVID epidemic in Northern Italy; we want instead to discuss -- with realistic parameters and framework -- what would be the differences if acting with different strategies in an epidemic with the general characteristics of the COVID one. Thus we will adopt the aforementioned parameters as ``bare'' ones (different strategies consisting indeed on acting on one or the other of these) but will apply these on a case study initial condition; this will be given by
\beql{eq:initial} I_0 \ = \ 10 \ , \ \ J_0 \ = \ 90 \ ; \ \ R_0 \ = \ U_0 \ = \ 0 \ . \eeq

One important parameter is missing from this list, i.e. the detection probability $\xi$. Following Li \emph{et al.} \cite{Li} we assumed in previous work that $\xi$ is between $1/10$ and $1/7$. Later works (and a general public interview by the Head of the Government agency handling the epidemic \cite{Bor}) suggested that the lower bound is nearer to the truth; moreover a lower $\xi$ will give us greater opportunity to improve things by acting on it (we will see this is not the best strategy, so it makes sense to consider the setting more favorable to it). We will thus run our simulation starting from a ``bare'' value $\xi = 1/10$.

As for the total population, we set $N=2*10^7$. With these choices we get
\beq \ga \ = \ 1.26 \ * \ 10^7 \ , \ \ \frac{S_0}{\ga} \ \simeq \ 1.58 \ . \eeq

We would like to stress once again that we will work with constant parameters, while in reality the parameters are changing all the time due to the continuing efforts to contain the epidemic. So our discussion is valid for what concerns the effect of different actions, but the absolute values of infected and other classes are by no means a forecast of what will happen; rather they should be seen -- in particular, those relating to the ``bare'' parameters -- as a projection of what could have happened if no action was undertaken.

\REM{6}{A note by an Oxford group \cite{Oxf}, much discussed (also in general press \cite{Guar}) upon its appearance, hinted that in Italy and UK this fraction could be as low as $\xi = 1/100$. We have ascertained that with this value of $\xi$, and assuming $\a$ was not changed by the restrictive measures adopted in the meanwhile, the A-SIR model fits quite well the epidemiological data available to the end of April. However, despite this, we do not trust this hypothesis -- at least for Italy -- for various reasons, such as (in order of increasing relevance): $(i)$ A viral infection showing effects only in 1\% of affected individuals would be rather exceptional; $(ii)$ Albeit in our opinion the effect of social distancing measures adopted in Italy is sometimes overestimated, we trust that there has been some effect; $(iii)$ if only 1\% of infected people was detected, in some parts of Italy the infected population would be over 100\%. On the other hand, the main point made by this report \cite{Oxf}, i.e. that only a large scale serological study, checking if people have COVID antibodies, will be able to tell how diffuse the infection is -- and should be performed as soon as possible -- is by all means true and correct. See also \cite{QMW}.}

\subsection{Balance between registered and unregistered infectives}

A look at eqs.\ref{eq:ASIR} shows that $I$ will grow provided \beq \frac{\xi \ S}{\ga} \ > \ \frac{I}{I+J} \ = \ \frac{I}{K} \ := \ x  \ , \eeq where again $\ga = \b / \a$, and we have introduced the ratio $x(t)$ of known infectives over total infectives. In other words, now the epidemic threshold \beq \ga_I \ = \ \( \frac{x}{\xi} \) \ \ga \eeq depends on the distribution of infectives in the classes $I$ and $J$. Note that if $x=\xi$ (as one would expect to happen in early stages of the epidemic), then $\ga_I = \ga$.

Needless to say, we have a similar result for $J$, i.e. $J$ will grow as far as  \beq (1 - \xi) \, S \ \frac{\a}{\eta} \ > \ \frac{J}{I+J} \ = \ \frac{J}{K} \ := \ y \ = \ 1 - x \ ; \eeq thus the epidemic threshold for unregistered infectives is \beq \ga_J \ = \ \( \frac{1-x}{1-\xi} \) \ \frac{\eta}{\a} \ . \eeq
For $x=\xi$ (see above) we would have $\ga_J = (\eta / \beta) \ga < \ga$.

It is important to note that $x$ is evolving in time. More precisely, by the equations for $I$ and $J$ we get
\begin{eqnarray} \frac{dx}{dt} &=& \a \, \xi \, S \ - \ \( \a S +\b - \eta \) \, x \ + \ (\b -\eta) \, x^2 \nonumber \\
&=& \a \, S \ ( \xi - x) \ + \ ( \b -\eta ) \, (x^2 - x) \ . \label{eq:xdot2} \end{eqnarray}

\begin{figure}
\centering
  \begin{tabular}{cc}
  \includegraphics[width=100pt]{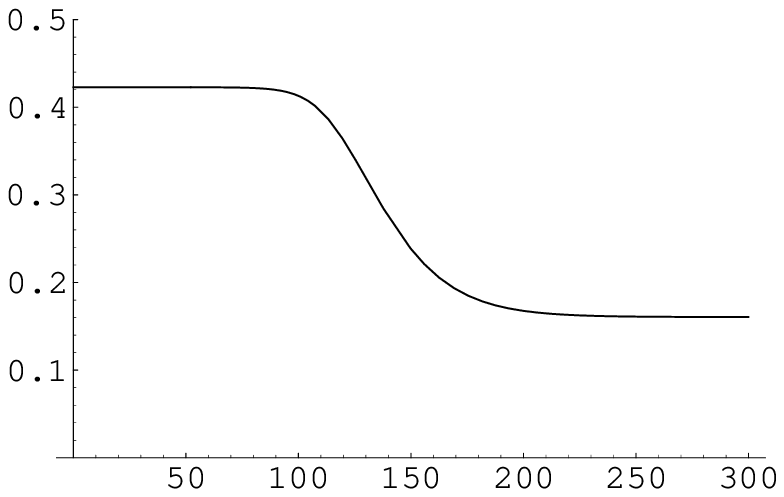} & \includegraphics[width=100pt]{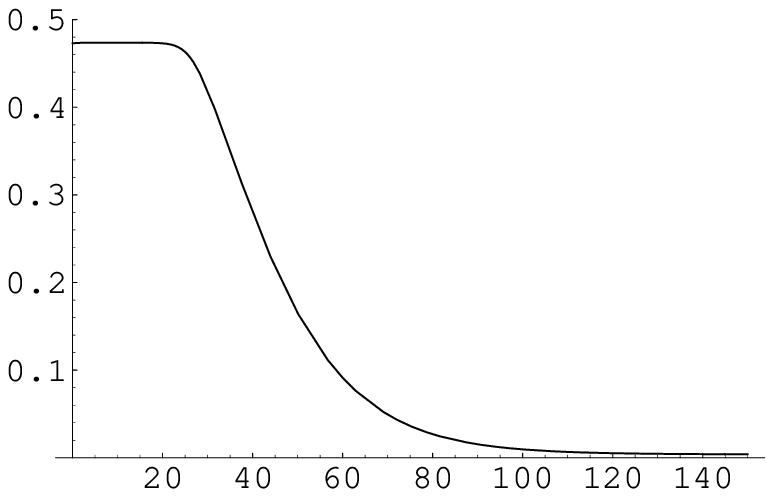} \\
  (a1) & (b1) \\
  \includegraphics[width=100pt]{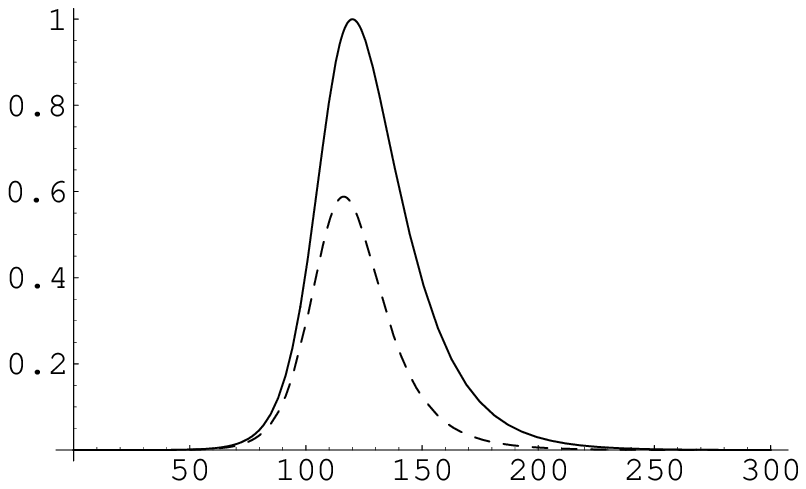} & \includegraphics[width=100pt]{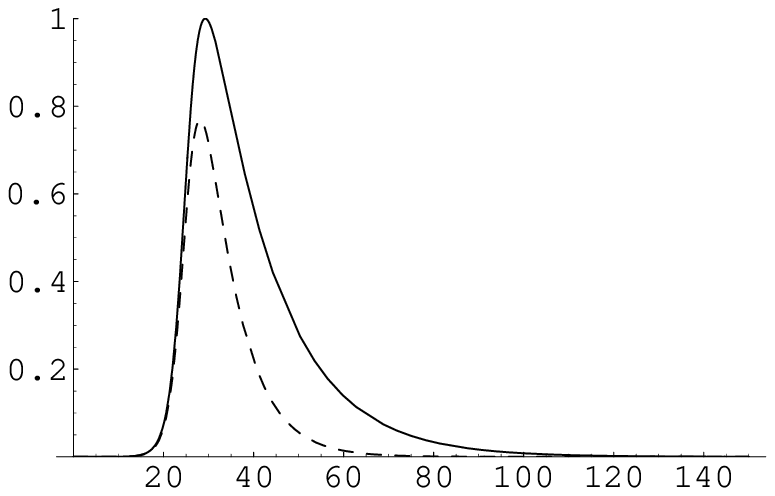} \\
  (a2) & (b2)
  \end{tabular}
  \caption{Dynamics of $x(t)$ in the A-SIR model. We plot $x(t)$ in upper plots (a1) and (b1); and $I(t)$ (dashed curve) and $J(t)$ (solid curve) in lower plots (a2) and (b2). These are considered along the solutions of the A-SIR model for $\a = 1.13 *10^{-8}$, $\beta = 1/7$, $\eta=1/14$, and $\xi = 1/10$; in the left column, i.e. in plots (a1) and (a2), with $S_0 = 2*10^7$ and $I_0+J_0 = 10$; in the right column, i.e. in plots (b1) and (b2), for $S_0 = 6*10^7$ and $I_0 + J_0 = 30$. The scale of plots (a2) and (b2) is chosen so that the maximum of $J(t)$ is at level one.}\label{fig:X}
\end{figure}

The behavior observed in Fig.\ref{fig:X}, which displays $x(t)$ and related quantities on a numerical solution of eqs.\eqref{eq:ASIR}, can be easily understood intuitively. In the first phase of the epidemic, there is an exponential growth of both $I$ and $J$; due to the structure of the equations, they grow with the same rate, so their ratio remains constant; on the other hand, once the dynamics get near to the epidemic peak, the difference in the permanence time of the two (that is, the time individuals remain in the infect class) becomes relevant, and we see (plots (a2) and (b2) of Fig.\ref{fig:X}) that not only the peak for $J$ is higher than the one for $I$, but it occurs at a slightly later time. Moreover, descending off the peak is also faster for $I$, as $\b^{-1} < \eta^{-1}$, and thus $x$ further decreases, until it reaches a new equilibrium while both classes $I$ and $J$ go exponentially to zero.

If we look at \eqref{eq:xdot2} we see that for fixed $S$ the variable $x$ would have two equilibria (one stable with $0 < x < 1$ and one unstable with $x>1$,stability following from $\b -\eta >0$), easily determined solving $dx/dt= 0$. Numerical simulations show that -- apart from an initial transient -- actually $x(t)$ stays near, but in general does not really sticks to, the stable fixed point determined in this way.

\subsection{The basic reproduction number}

A relevant point should be noted here. If we consider the sum \beq K(t) \ := \ I(t) \ + \ J(t) \eeq of all infectives, the A-SIR model can be cast as a SIR model in terms of $S$, $K$, and $Q = R+U$ as
\begin{eqnarray}
dS/dt &=& - \, \a \, S \, K \nonumber \\
dK/dt &=& \a \, S \, K \ - \ B \, K \\
dQ/dt &=& B \, K \nonumber \end{eqnarray}
where $B$ is the \emph{average} removal rate, i.e.
$$ B \ = \ x \, \b \ + \ (1 -x) \, \eta \ . $$
As $x$ varies in time, this average removal rate is also changing. On the other hand, the \emph{basic reproduction number} (BRN) $\rho_0$ (this is usually denoted as $R_0$, but we prefer to change this notation in order to avoid any confusion with initial data for the known removed $R(t)$) for this model will be
\beql{eq:rho0} \rho_0 \ = \ \frac{\a}{B} \ S \ > \ \frac{\a}{\b} \ S \ . \eeq
In other words, not taking the asymptomatic infectives into account leads to an \emph{underestimation} of the BRN. If the standard SIR model predicts a BRN of $\rho_0$, the A-SIR model yields a BRN $\^\rho_0$ given by
\beq \^\rho_0 \ = \ \frac{\b}{B} \ \rho_0 \ = \ \frac{\b}{x \b + (1-x) \eta} \ \rho_0 \ > \ \rho_0 \ . \eeq

This means that the epidemic will develop faster, and possibly much faster, than what one would expect on the basis of an estimate of $\rho_0$ based only on registered cases, which in the initial phase are a subset of symptomatic cases as the symptoms may easily be leading to a wrong diagnosis (in the case of COVID they lead to a diagnosis of standard flu).

With our COVID-related values $\b = 1/7$, $\eta = 1/21$, and assuming that in the early phase $x = \xi = 1/10$, we get
\beq \^\rho_0 \ = \ \frac{5}{2} \ \rho_0 \ ; \eeq
there is thus a good reason for being surprised by the fast development of the epidemic: \emph{the actual BRN is substantially higher than the one estimated by symptomatic infections} \cite{GBRN}.

\section{Hidden infectives and epidemic dynamics}

More generally, one would wonder what is the effect of the ``hidden'' infectives $J(t)$ on the dynamics of the known infectives $I(t)$ -- which, we recall, include the relevant class of seriously affected infectives -- and it appears that there are at least two, contrasting, effects:

\begin{enumerate}

\item On the one hand, the hidden infectives speed up the contagion spread and hence the rise of $I(t)$;

\item On the other hand, they contribute to group immunity, so the larger this class the faster (and the lower the $I$ level at which) the group immunity will be reached.

\end{enumerate}

The discussion above shows that the balance of these two factors leads to a much lower epidemic peak, and a shorter epidemic time, than those expected on the basis of the standard SIR model (albeit in the case of COVID with no intervention these are still awful numbers).

On the other hand, we would like to understand if uncovering a larger number of cases (thus having prompt isolation of a larger fraction of the infectives) by \emph{early detection}, i.e. raising $\xi$, would alter the time-span of the epidemic. It appears that this effect can be only marginal, as it appears only past the epidemic peak.

We stress that this statement refers to ``after incubation'' analysis; if we were able to isolate cases \emph{before} they test positive -- i.e. to substantially reduce $\b^{-1}$ -- the effect could be different. We will discuss this point, related to \emph{contact tracing}, later on.

\subsection{COVID. Observable and ``clean'' observable data}

An ongoing epidemic is not a laboratory experiment, and apart from not having controlled external conditions, i.e. constant parameters, the very collection of data is of course not the top priority of doctors fighting to save human lives.

There has been considerable debate on what would be the most reliable indicator to overcome at least the second of these problems. One suggestion is to focus on the number of deaths; but this is itself not reliable, as in many cases COVID is lethal on individuals which already had some medical problem, and registering these deaths as due to COVID or to some other cause depends on the protocol  adopted, and in some case also on political choices, e.g in order to reassure citizens (or on the other extreme, to stress great care must be taken to avoid contagion).

Another proposed indicator, possibly the most reliable in order to monitor the development of the epidemic, is that of patients in Intensive Care Units. This appears to be sufficiently stable over different countries, and e.g. the Italian data tend to reproduce in this respect - at least in Regions where the sanitary system is not overstretched -- the Chinese ones.

In this case, IC patients are about 20 \% of the total number of hospitalized cases; in China and for a long time also in Italy (when protocols for choosing would-be cases to be subject to laboratory analysis have been stable), hospitalized cases have been about half of the known infection cases, the other having shown only minor symptoms and been cured (and isolated) in their home.

The other, more widely used, indicator is simply the total number of known cases of infection. In view of the presence of a large class of asymptomatic infectives, this itself is strongly depending on the protocols for chasing infectives. On the other hand, this is the most available indicator: e.g., the W.H.O. situation reports \cite{WHOrep} provide these data.

Each of these indicators, thus, has advantages and disadvantages. We will just use the WHO data on known infected.

In particular, in the case of COVID we expect that with $\xi_0$ the ``bare'' constant describing the probability that an infection is detected, out of the class $I(t)$ we will have a 50\% of infected with little or no symptoms $(I_L$), a 40\% of standard care hospitalized infected ($I_H$), and a 10\% of IC hospitalized infected ($I_{IC}$). Needless to say, this class is the most critical one, also in terms of strain on the health system.

More generally, we say that with $\xi_0$ the ``bare'' constant describing the probability that the infection under study is detected, there is a fraction $\chi_0$ (of the detected infections) belonging to the $I_{IC}$ class; that is, $I_{IC} (t) = \chi_0 I (t)$.

\REM{7}{We stress this depends on the protocol used to trigger laboratory tests; in our general theoretical discussion, this is any such protocol and we want to discuss the consequences of changing this in the sense of more extensive tests.}

\section{Modifying the parameters}

We are now ready to discuss how modification of one or the other of the different parameters ($\a,\b,\xi$) on which we can act by various means will affect the A-SIR dynamics. As it should be expected, this will give results similar to those holding for the SIR model, but now we have one more parameter to be considered and thus a more rich set of possible actions.

\subsection{Raising the detected fraction}

\begin{figure}
\centering
  \includegraphics[width=200pt]{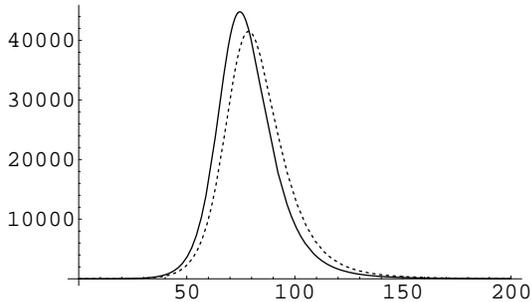}\\
  \caption{The effect of a change in $\xi$ on the $I_{IC}$ class. We have used $\b=1/7$, $\eta=1/21$, and $\a = 1.13 *10^{-8}$ as in Fig.2, with a total population of $N=2*10^7$, and ran simulations with $\xi = 1/10$ (solid curve) and with $\xi = 1/4$ (dashed curve). The substantial increase in $\xi$ produces a reduction in the epidemic peak and a general slowing down of the dynamics, but both these effects are rather small.}\label{fig:xi}
\end{figure}

A more extensive test campaign will raise $\xi$, say from $\xi_0$ to $\xi_1$; but of course this will not change the number of the most serious cases, as these are anyway getting to hospital and detected as being due to the infection in question. Thus the new fraction $\chi_1$ of detected infections which need special care will be such that $\chi_1 \xi_1 = \chi_0 \xi_0 $, i.e. we have
\beql{eq:chi} \chi_1 \ = \ \frac{\xi_0}{\xi_1} \ \chi_0 \ . \eeq
In order to describe the result of raising $\xi$, we should thus compare plots of \beql{eq:IIC} I_{IC} (t) \ = \ \chi \ I (t) \ . \eeq
This is what we do, indeed, in Fig.\ref{fig:xi}.

\subsection{Running ahead of the epidemic wave}

Raising $\xi$ corresponds to having more infective detected, and has some advantages from the point of view of the epidemic dynamics. In practical terms, this means extending tests to a larger class of subjects, and be able to isolate a larger fraction of asymptomatic infectives with the same speed and effectiveness as symptomatic ones.

A different strategy for rapid action is also possible, and it consists of rapid isolations of subjects who had contacts with people known to have been infected, or who have themselves been in contact with known infectives (and so on). In other words, the strategy would be to isolate would-be infection carriers \emph{before} any symptom could show up. This means that $\b^{-1}$ could be even smaller than the usual infection-to-isolation time (about seven days for COVID) for symptomatic infectives, and even shorter than the incubation time (about five days for COVID).

\REM{8}{It should be stressed that as each of these ``possible infected'' might have a small probability of being actually infected (depending on the kind of contacts chain leading to him/her from known infectives), here ``isolation'' does not necessarily mean top grade isolation, but might amount to a very conservative lifestyle, also -- and actually, especially -- within home, where a large part of registered Chinese contagions took place. {(The same large role of in-home contagion was observed in Italy in the course of lockdown.)}} \bigskip

\begin{figure}
\centering
  \includegraphics[width=200pt]{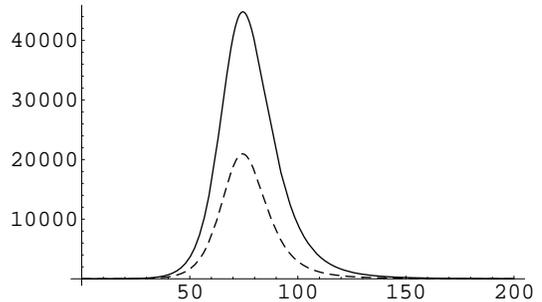}\\
  \caption{The effect of a change in $\b$ on the $I_{IC}$ class. We have used $\xi=1/10$, $\eta=1/21$, and $\a = 1.13 *10^{-8}$ as in Fig.2, with a total population of $N=2*10^7$, and ran simulations with $\b = 1/7$ (solid curve) and with $\b = 1/3$ (dashed curve). The substantial increase in $\b$ produces a marked reduction in the epidemic peak and a very slightly faster pace in the dynamics.}\label{fig:beta}
\end{figure}

We have thus ran a simulation in which $\xi$ is not changed, but $\b$ is raised  from $\b_0 = 1/7$ to $\b = 1/3$; the result of this is shown in Fig.\ref{fig:beta}.
In this case we have a marked diminution of the epidemic peak, and a very slight acceleration of the dynamics.

\subsection{Social distancing}

We have so far not discussed the most basic tool in epidemic containment, i.e. social distancing. This means acting on the parameter $\a$ by reducing it.

\REM{9}{Direct measurement on the epidemiological data for Northern Italy show that this parameter can be reduced to about 20\% of its initial value with relatively mild measures. In fact, albeit the media speak of a generalized lockdown in Italy, the measures have closed schools and a number of commercial activities, but for the rest were actually more pointing at limiting leisure walk and sports and somewhat avoiding contacts in shops or in work environment than to a real lockdown as it was adopted in Wuhan.} \bigskip

This is a basic action to be undertaken, and in fact it is being taken by all Nations. It is also the simplest one to be organized (albeit with high economic and social costs in the long run) and an action which can be taken together with other ones. No doubt this should be immediately taken when an epidemic is starting, and accompanied by other measures -- such as those discussed above. But here we want to continue our study of what it means by itself in terms of modification of the epidemic dynamics.

It is not clear what can be achieved in terms of reduction of social contacts. In fact, once the epidemic starts most of the dangerous contacts are the unavoidable ones, such as those arising from essential services and production activity (e.g. production and distribution of food or pharmaceutical goods), contacts at home, and above all contacts in Hospitals. Thus, after a first big leap downward corresponding to closing of schools and Universities on the one side, and a number of unessential commercial activities on the other, and restrictions on travels, it is difficult to further reduce social contacts, not to say that this would have huge economic and social costs, and also a large impact on the general health in terms of sedentariness-related illness (and possibly mental health).

\REM{10}{A number of countries tried to further reduce social contacts by forbidding citizens to get out of their home; this makes good sense in densely populated areas, but is useless in many other areas. The fortunate slogan ``stay home'' risks to hide to the general public that the problem is not to seclude oneself in self-punishment, but to \emph{avoid contacts}.} \bigskip

\subsection{Social distancing and epidemic timescale}

We point out that there is a further obstacle to reducing social contacts: as seen in the context of the simple SIR model, reducing $\a$ will lower the epidemic peak, but it will also \emph{slow down the whole dynamic}. While this allows to gain precious time to prepare Hospitals to stand the big wave, there is some temporal limit to an extended lockdown, and thus this tool cannot be used to too large an extent.

\begin{figure}
\centering
  \includegraphics[width=200pt]{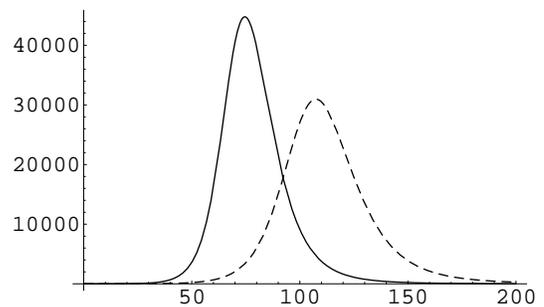}\\
  \caption{The effect of a change in $\a$ on the $I_{IC}$ class. We have used $\b=1/7$, $\xi=1/10$, $\eta=1/21$, with a total population of $N=2*10^7$, and ran simulations with $\a = 1.13 *10^{-8}$ (solid curve) and with $\a = 8.47 *10^{-9}$ (dashed curve). The reduction in $\a$ produces a marked reduction in the epidemic peak and also a marked slowing down in the dynamics.}\label{fig:ACO}
\end{figure}

We have thus ran a simulation in which $\b$ and $\xi$ are not changed, while $\a$ is reduced by a factor $0.75$ (smaller factors, i.e. smaller$\a$, produce an untenable length of the critical phase); the result of this is shown in Fig.\ref{fig:ACO}.
In this case we have a relevant diminution of the epidemic peak, and also a marked slowing down in the dynamics.

An important remark is needed here. It may seem, looking at this plot, that social distancing is less effective than other way of coping with the epidemic. But these simulation concern a SIR-type model; this means in particular that there is no spatial structure in our model \cite{Murray}. The travel ban is the most effective way of avoiding the spreading of contagion from one region to the others; while the ``local'' measures of social distancing can (and should) be triggered to find a balance with other needs, travel ban is the simplest and most effective way of protecting the communities which have not yet been touched by the epidemic.

\subsection{Comparing different strategies}

\begin{figure}
\centering
  % Requires \usepackage{graphicx}
  \includegraphics[width=200pt]{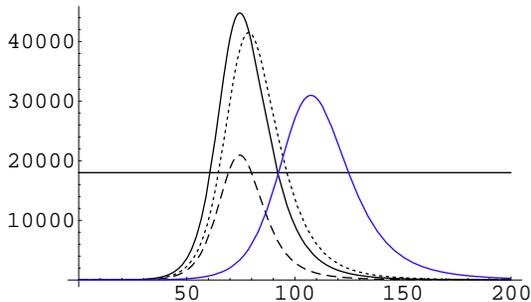}\\
  \caption{The effect of different strategies. We plot $I_{IC} (t)$ for $N=2*10^7$ in the ''bare'' case, i.e. for $\a = 1.13 *10^{-8}$, $\b=1/7$, $\xi=1/10$, $\eta=1/21$, and in cases where (only) one of the parameters is changed. In particular we have the bare case (solid line), the case where $\xi$ is changed into $\xi=1/4$ (dotted), the case where $\b$ is changed to $\b = 1/3$ (dashed), and that where $\a$ is changed to $\a = 8.47 *10^{-9}$ (solid, blue). We also plot a horizontal line representing a hypothetical maximal capacity of IC units.}\label{fig:all}
\end{figure}

We can thus compare the different strategies we have been considering. This is done in Fig.\ref{fig:all} where we plot together $I_{IC} (t)$ for all our different simulations; and in Table \ref{tab:all} where we compare the height of the epidemic peak -- again for $I_{IC} (t)$ -- and the time at which it is reached.

In Fig.\ref{fig:all} we have also drawn a line representing the hypothetical maximal capacity of IC units. This stresses that not only the different actions lower the epidemic peak, but they also -- and to an even larger extent -- reduce the number of patients which can not be conveniently treated.

\REM{11}{In looking at this plot, one should remember that the model does not really discuss permanence in IC units, and that $I_{IC}$ are the infected which \emph{when detected} will require IC treatment; this may go on for a long time -- which is the reason why IC units are saturated in treating COVID patients. So the plots are purely indicative, and a more detailed analysis (also with real parameters) would be needed to estimate the IC needs in the different scenarios.}
\bigskip

It should be stressed that the strategies of contacts tracing and early detection are usually played together; but as confusion could arise on this point, let us briefly discuss it. We have tried to stress that these two actions are \emph{not} equivalent: one could conduct random testing, so uncovering a number of asymptomatic infectives, and just promptly isolate them without tracing their contacts;or on the other extreme one could just isolate everybody who had a (direct or indirect) contact with a known infective, without bothering to ascertain if they are themselves infective or not. This strategy would be as effective in containing the contagion (and less costly in terms of laboratory tests) than that of tracking contacts, test them (after a suitable time for the infection to develop and test give positive if this happens), and isolate only those who really turn infective. The difference is that if we isolate everybody this would involve a huge number of people (e.g. all those who have been in the same supermarket the same day as an infective; and their families and contacts etc etc); so in this context early detection actually should be intended as early detection \emph{of non-infectives}, so that cautionary quarantine can be kept reasonably short in all the cases where it is not really needed.

Finally we recall that it is a triviality, and it was already mentioned in the Introduction, that in real situations one has not to choose between acting on one or the other of the parameters, and all kind of actions should be pursued simultaneously.

\begin{table}
  \centering
  \begin{tabular}{|c|c|c||r|r|}
  \hline
  $\a$ & $\b$ & $\xi$ & max & time \\ % $I_{IC}$
  \hline
  $1.13 *10^{-8}$ & 1/7 & 1/10 & 44768 &  75 \\
  $1.13 *10^{-8}$ & 1/7 & 1/4  & 41482 &  79 \\
  $1.13 *10^{-8}$ & 1/3 & 1/10 & 20943 &  74 \\
  $8.47 *10^{-9}$ & 1/7 & 1/10 & 30956 & 107 \\
  \hline
  \end{tabular}
  \caption{Epidemic peak (for $I_{IC}$) and time for reaching it (in days) as observed in our numerical simulations. All simulation were ran with $N=2*10^7$ and $\eta=1/21$.}\label{tab:all}
\end{table}

\subsection{A different indicator}
\label{sec:mu}

The numerical computations of the previous subsections suggest that increasing $\xi$ -- that is, detection of a larger fraction of asymptomatic -- is not a very efficient strategy to counter the diffusion of an infection with a large number of asymptomatic infectives, while a prompt isolation of infectives is a more effective action.

It should be recalled, however, that in our computations -- and in particular on Figure \ref{fig:all}, where their outcomes are compared -- we are focusing on the number of patients needing IC support, i.e. the most critical parameter from the point of view of the Health system. In order to substantiate our conclusions, it is worth considering also different ways to evaluate the effect of different strategies.

We have thus considered also a different indicator, i.e. the total number of infectives $$ K(t) \ := \ I(t) \ + \ J(t) \ . $$ We have run several simulations, with total population $N= 2*10^7$ and with parameters
\beql{eq:mu1} \a = \mu_\a \a_0 \ , \ \ \b = \ \mu_\b \b_0 \ , \ \ \eta = \mu_\eta \eta_0 \ , \ \ \xi = \mu_\xi \xi_0 \ . \eeq Here $\mu_i$ are modulation factors describing the changes in the basic parameters, and
\beql{eq:mu2}\a_0 = 1.13 * 10^{-8} \ , \ \ \b_0 = 1/7 \ , \ \ \eta_0 = 1/21 \ , \ \ \xi_0 = 1/10 \eeq are reference values (already used in the previous subsections). The initial conditions where chosen so that \beql{eq:mu3} K_0 = 100 \ ; \ \ I_0 = \xi K_0 \ , \ \ J_0 = (1-\xi) K_0 \ . \eeq

\begin{figure}
  \centering
  \includegraphics[width=200pt]{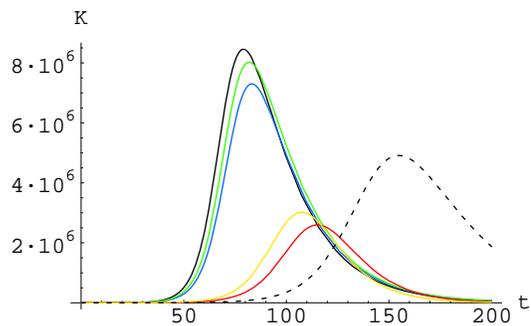}\\
  \caption{Plot of the total number of infectives $K(t) = I(t) + J(t)$ for parameters as in \eqref{eq:mu1}, \eqref{eq:mu2} and with different choices of the modulation factors $\mu_i$. The initial conditions are chosen according to \eqref{eq:mu3}. In the reference run (black curve, solid) all the $\mu_i$ are equal to 1; in the other runs the modified factors are as follows: Change of $\xi$ (blue), $r_\xi = 2.5$; change of $\b$ alone (green), $r_\b = 7/3$; change of $\eta$ alone (yellow), $r_\eta = 7/3$; change of both $\b$ and $\eta$ (red), $r_\b = r_\eta = 7/3$; change of $\a$ (black, dashed), $r_\a = 3/5$. }\label{fig:mu}
\end{figure}

The outcome of these simulations is displayed in Figure \ref{fig:mu}; see its caption for the parameter (that is, the modulation factor) values in different runs.

We see from Figure \ref{fig:mu} that action on $\a$ slows down substantially the epidemic dynamics\footnote{The factor $r_\a = 3/5$ was chosen so not to have it completely off-scale with respect to the reference run.} and reduce the epidemic peak, while action on $\xi$ or on $\b$ alone produce only a moderate effect. On the other hand, actions affecting the value of $\eta$ (alone or together with the value of $\b$) reduce substantially the epidemic peak and slightly slow down the dynamics.

\REM{12}{It may be noted that the shapes of the $I_{IC} (t)$ (see Figure \ref{fig:all}) and of the $K(t)$ (see Figure \ref{fig:mu}) are different; in particular, the decay of $I_{IC} (t)$ after attaining its peak is faster than the decay of $K(t)$. This corresponds to what is observed in the epidemiological data for Italy.}

\section{Discussion and conclusions}

We have considered epidemic dynamics as described by ``mean field'' models of the SIR type; more specifically, we have first considered the classical Kermack-McKendrick SIR model \cite{KMK,Murray,Britton,Heth,Edel} and then a recently introduced modified version of it \cite{Gasir} taking into account the presence of a large set of asymptomatic -- and thus most frequently not detected -- infectives. These models depend on several parameters, and different types of measures can to some extent change these parameters and thus the epidemic dynamics. In particular, this action can effect two basic characteristics of it, i.e. the height of the epidemic peak and the time-span of the epidemic.

While it is clear that in facing a real lethal epidemics (such as the ongoing COVID epidemic) all actions which can contrast it should be developed at the same time, in this paper we have considered the result -- within these models -- of different tools at our disposal, i.e. (generalized) \emph{social distancing}, \emph{early detection} (of asymptomatic infectives) and \emph{contacts tracing} (of symptomatic and asymptomatic infectives).

It turns out that -- both in the classical SIR model and in the modified A-SIR one -- social distancing is effective in reducing the epidemic peak, and moreover it slows down the epidemic dynamics. On the other hand, \emph{early detection} of asymptomatic infectives seems to have only a moderate effect in the reduction of the epidemic peak for what concerns critical cases, and also a very little effect on the temporal development of the epidemic. In contrast, \emph{contact tracing} has a strong impact on the epidemic peak -- also in terms of critical cases -- and does not substantially alter the temporal development of the epidemic, { at least for what concerns the curve describing the most serious cases}.\footnote{One point needs maybe further discussion. A tight social distancing policy is, after all, keeping people from having contacts and is thus equivalent to isolate not only would-be infectives but everybody; thus this should be equally effective. The point is that a ``no contacts for everybody'' policy is simply not feasible: ill people need help, people living in cities need to buy food, a number of essential services simply cannot be stopped.}

\REM{13}{The conclusion that early detection of asymptomatic has only a moderate effect may appear to be paradoxical, and requires some further discussion. First of all we should remind that we are here actually talking about an increase of the parameter $\xi$ (see Remark 4), while in a real situation early detection of asymptomatic will most likely go together with early detection of symptomatic, and hence a reduction in $\b$ as well. The increase of $\xi$ \emph{per se} means that some fraction of asymptomatic will be recognized as infective and be isolated on the same timescale $\b^{-1}$ as the symptomatic infectives, while the other asymptomatic will escape recognition and still be infective on a timescale $\eta^{-1}$. On the other hand, a realistic \emph{contact tracing} campaign will lead to prompt isolation of symptomatic and asymptomatic alike, and thus correspond to a reduction in $\b^{-1}$ and in $\eta^{-1}$, and we have seen that this action is indeed the most effective one in terms of contrasting the spread of the epidemic.
In other words, our result suggests that the key to fight COVID is not so much in \emph{detection}, but in \emph{prompt isolation} of infectives, and most notably of asymptomatic ones. This can be achieved only by \emph{contact tracing} -- as already suggested by experienced epidemiologists.\footnote{To give further strength to this conclusion, note that if we run a simulation with the same parameters as in Section \ref{sec:mu} above but with $\eta = \beta = 1/7$ -- i.e. assuming asymptomatic are isolated after the same characteristic time as symptomatic from infection -- what is observed is that $K$ is steadily decreasing.}}
\bigskip

Slowing down the epidemic dynamic can be a positive or negative feature depending on the concrete situation and on the desired effects. It is surely positive in what concerns getting ready to face the epidemic peak, in particular in the presence of a faltering Health System. On the other hand, it may be negative in that maintaining a generalized lockdown for a long time can have extremely serious economic and social consequences. Balancing these two aspects is not a matter for the mathematician or the scientist, but for the decision maker; so we will not comment any further about this.

It should also be recalled that our analysis was conducted in terms of very simple SIR-type models, with all their limitations. In particular, we have considered no age or geographical or social structure, and only considered a population of ``equivalent'' individuals. In particular, as we have noted above, in the early stage of an epidemic, which presumably develops in very populated areas, a generalized travel ban can simply stop the contagion to propagate to other (possibly less well equipped in medical terms) areas; moreover, social distancing measures can be implemented very simply -- basically, by a Government order (albeit if we look at the goal of these measures, i.e. reducing the occasion of exchanging the virus, a substantial role would be played by individual protection devices, such as facial masks; in many European countries, these were simply not available to the general public, and in some cases neither to medical operators, thus substantially reducing the impact of these measures) -- and are thus the first action to be taken. In fact, in relation with the ongoing COVID epidemics, one of the reproaches made to many Governments is usually to have been too slow or too soft in stopping crowd gatherings, surely not the contrary.

On the other hand, we hope that this study makes clear what are the consequences of different options. In particular, our study shows that \emph{contacts tracing},  followed by prompt isolation of would-be infected people -- is the only way to reduce the impact of the epidemic without having to live with it for an exceedingly long time. The Veneto experience \cite{Cri} shows that this strategy can be effectively implemented without hurting privacy or personal freedom.

%\newpage

\section*{Acknowledgements}

The work was carried out in lockdown at SMRI. I am also a member of GNFM-INdAM.

%\newpage

\begin{appendix}

\section{An aftersight on the A-SIR  model}

Our discussion was based on SIR-type models, and in particular on the A-SIR model. This raises several kind of questions, which we address in this Appendix.

\begin{enumerate}

\item The choice of dealing with SIR-type (i.e. compartment) models in dealing with the COVID-19 epidemic could be questioned.

\item Within this class, the A-SIR model is specially simple; it makes sense to rely on this model only if it is able to give a reasonable agreement with observed data.

\item The A-SIR model is the simplest one taking into account the presence of asymptomatic infectives. Other -- more detailed -- models taking this fact into account have also been developed.

\end{enumerate}

We are now going to briefly discuss these matters; we point out that this Appendix was inserted in the revised version of this paper, so it can make use of knowledge not available at the time of writing the first submitted version, nd mentions papers appeared after the first submittal.

\subsection{Use of SIR-type models}

Compartment models, i.e. SIR-type ones in this context, are based on several implicit and explicit assumptions, which are not realistic in many cases and surely when attempting to describe the COVID epidemic, in particular in a full country.

That is, among other aspects, SIR-type models are (in Physics' language) \emph{mean field (averaged) models} and as such describe the dynamics and the underlying system as if:

\begin{itemize}
\item All individuals are equivalent in medical sense, i.e. they all have equivalent pre-existent health status and equivalent immune system and react in the same way to contact with the pathogen;
\item In particular, as we know that COVID is statistically more dangerous for older people, we are completely disregarding the \emph{age structure} of the population, as well as the existence of other high risk classes related to pre-existent pathologies: all these contribute to an average over the whole population;
\item All individuals are equivalent in social sense, i.e. they all have an equivalent social activity and hence the same number and intensity of contacts with other members of the group, thus the same exposure to (possible) infectives;
\item In particular, this means we are completely disregarding any \emph{geographical structure} in the population, and consider in the same way people living in large cities or in remote villages, just considering them in the same global average;
\item Similarly, we do not consider that work can cause some people to be specially exposed through contact with a large number of people (e.g. shop cashiers) or even with a large number of infected people (e.g. medical doctors or nurses).
\end{itemize}

Thus one cannot hope to retain, through such models, effects like the faster spreading of the infection in more densely populated areas or the specially serious consequences of the COVID infection among older people.

We stress that these could be obtained by including geographical, age or social structures into the model, i.e. increasing the number of considered classes; in principles this should provide a finer and more realistic description of the epidemic dynamic, and in fact it is done in cases for which there is a large set of data, e.g. for influenza. Such structured models would of course loose the main attractive of the SIR model, i.e. its simplicity -- which also allows to understand in qualitative terms the mechanisms at work.

In particular, a relevant intermediate class of models is that of SIR-type models on networks: these take into account geographical and social structures and make use of known information about contacts between different groups of individuals and about different health characteristics of different groups.

The problem with these networked or however more structured models is that the network should be inferred from data. In this respect, it could be objected that the influenza monitoring over many years could give us the relevant data for reconstruction of such network; but it is everybody's experience, by now, that the social behavior of people are completely different if dealing with a well known and not so serious (except for certain categories) illness like influenza or with an unknown and potentially lethal one like COVID; this not to say that the restrictive measures put into effect by many Governments have completely changed the interaction patterns among people, so that previously accumulated data cannot be used in the present situation.

When thinking of COVID; it should be kept in mind that even if the countries which were first hit by the epidemic, we only have data over some months; e.g. for Italy we have about 100 days of data. If we were trying to give the model a geographical structure at the level of Departments (which are themselves administrative units mostly with a very varied internal geographical structure), as there are 107 Departments in Italy this would require in the simplest form to evaluate a $107 \times 107$ interaction matrix, and I cannot see any way to reliably build this out of such a scarce set of data.

Moreover, the epidemiological data are to some extent not reliable, especially around the epidemic peak, in that they are collected in an emergency situation, when other priorities are present in Hospitals (e.g. in Italy the data show a weekly modulation, which appears to be due simply to the procedure of data collection); so an even larger amount of data would be needed to filter out statistical noise and random fluctuations.

In this sense, the weak point of SIR-type models, i.e. their being based on an average over the whole population, turns out to be an advantage: they contain few parameters (two for the SIR, four for the A-SIR) and are thus statistically more robust in that fluctuations are averaged efficiently with less data than for more refined models with a large number of parameters.

\REM{A1}{Similar considerations hold when one compares SIR-type models to a purely statistical description or to an ``emerging behavior'' approach. These approaches are extremely powerful, but are effective when one has a large database to build on and to which compare the outcome of the ``experiment'' (in this case the epidemic) under consideration. When we deal with a completely new pathogen,like for COVID, we simply don't have a database, and we can only rely on the very general features of infective dynamics -- which are well coded by SIR and SIR-like models.}
\bigskip

In other words, we are not claiming the SIR approach to be superior to others, but only that it is appropriate when we have few data -- as for COVID.

\subsection{Use of the A-SIR model}

Within the SIR-type class, the A-SIR model is specially simple; from the theoretical point of view its appeal lies in that it is the simplest possible model taking into account the presence of a large class of asymptomatic infectives; thus it focuses on the effect of this fact without the complications of a more detailed model. But, of course, it makes sense to rely on this model only if it is able to give a good, or at least a reasonable, agreement with observed data.

Of course each infective agent has its own characteristics, and using only the general SIR model would completely overlook them, apart from the different values of the $\a$ and $\b$ parameters.

Thus we have to do something more than just evaluating the SIR parameters. In our study we have identified the presence of a large class of asymptomatic infectives  as one of the key problems in facing the COVID epidemic, and we have considered a simple model which allows to focus precisely on this aspect.

One should be aware that the A-SIR model is focusing on this and not considering other features of COVID, and indeed other more detailed SIR-type models for COVID have been formulated and studied (see also below). Here we are taking an approach which is classical in Mathematical Physics and Mathematical Modeling, i.e. try to build and study the simplest model describing the phenomenon of interest. This will give results which are quantitatively worse than a more detailed models, but which are qualitatively good in that the model is simple enough to see more clearly what are the mechanisms at work and to understand the qualitative features if the dynamics and the qualitative outcome of any intervention able to modify the parameters of the model.

Having said that, it remains true that -- as mentioned above -- it makes sense to rely on this model only if it is able to give a reasonable agreement with observed data. This is not the argument of this paper, and it was discussed in a previous paper \cite{Gasir}; the success of this model was the justification for this paper,i.e. for discussing the effect of different COVID-contrasting strategies in terms of it.

However, the first version of this paper was submitted at mid-May, hence with two and half months of data available, while at the time of preparing this revised version we have four months of data; that represents a substantial increase in the available data, and it makes sense to wonder if the model is still describing the COVID epidemic in Italy.

\begin{figure}
  \centering
  \includegraphics[width=200pt]{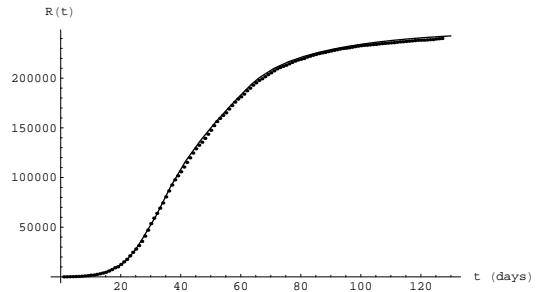}\\
  \caption{Epidemiological data against solutions of the A-SIR equations \eqref{eq:ASIR} for Italy. Time is measured in days, with day 0 being February 20, 2020. See text for parameter values and initial conditions.}\label{fig:R}
\end{figure}

This is indeed the case, as shown in Figure \ref{fig:R}; they represent epidemiological data as communicated by the Italian Health Ministry and by WHO (and widely available online through the standard COVID databases) against a numerical integration of the A-SIR equations \eqref{eq:ASIR}. We refer to \cite{Gasir} for a discussion of the parameters and their determination. Note that the contact rate $\a$ is assumed to vary in response to the restrictive measures (and to the availability of individual protection devices); as these measures were taken in different steps, we also have different values of $\a$ in different time intervals.

More precisely, the equations were integrated for a total population of $N=6*10^7$ for the period February 20 through June 26 with initial data at day 14 (March 5) given by $I_0 = 6766$, $J_0 = 60892$, $R_0 = 3862$, $U_0 = 34761$, with  $\a = r(t) \a_0$ where
$$ r(t) \ = \ \cases{1 & for $t \le 25$ \cr 0.5 & for $25 < t \le 35$ \cr 0.2 & for $35 < t \le 63$ \cr 0.08 & for $63 < t$ \cr} $$ and the parameters are given by
$$ \a_0 \simeq 3.77*10^{-9} \ , \ \ \b = 1/7 \ , \ \ \eta = 1/21 \ , \ \ \xi = 1/10 \ . $$
We stress that the parameter values are the same as in \cite{Gasir}, even for the most recent time, not considered in that paper: the model continues to reasonably well describe the development of the epidemic in Italy.

\subsection{Other SIR-type models}

We focused on a specific SIR-type model, but several models of this type have been considered in the context of COVID modeling. Here we give a very short overview of these, with no attempt to completeness -- which cannot even be imagined in such a rapidly evolving field.

First of all, we note that other researchers have considered, motivated by the ongoing COVID epidemic, the temporal aspects of the standard SIR dynamics. We mention in particular Cadoni \cite{Cadoni} (a related, but quite involved, approach had been considered by Harko, Lobo and Mak \cite{HLM}) and Barlow and Weinstein \cite{BaW}, who obtained an exact solution for the SIR equations in terms of a divergent but asymptotic series \cite{Ba17}; see also \cite{BBGS,Gra} for a different approach to exact solution of SIR and SIR-type models.

We also note that nonlinear modifications of the bilinear infection
term of the standard SIR model have been proposed -- explicitly or implicitly -- in the attempt to relate the standard SIR model to COVID dynamics \cite{secord,Volpert}. We find \cite{Volpert} of special interest, as this work introduces a model for the epidemic dynamics coupled to the immune system, and is thus able to take into account aspects related to the \emph{viral charge} of infectives.

Extension of the SIR model in the direction of allowing time-dependence of the parameters -- also to account for shifting public attitude -- has also been considered \cite{CLCL}.

As mentioned above, see Remark 3, considering the delay between infection and beginning of infectiveness would lead to consider SEIR-type models. The problem of temporal aspects of the dynamics for this class of models has been considered by Becaer \cite{Becaer}. The role of asymptomatic transmission in this class of models has also been considered \cite{Arcede,Picchiotti}. The approach to SIR by Barlow and Weinstein \cite{BaW} leading to exact solution has been extended to SEIR model \cite{BaW2}.

A generalization of the A-SIR model, allowing for different infectiveness of symptomatic and asymptomatic infectives, has been considered by Neves and Guerrero \cite{NeGue}.

More elaborated compartment models with a larger number of compartments have been considered by a number of authors. We would like to mention in particular two papers which we consider specially significant, i.e. the work by the Pavia group, in which mathematicians, statisticians and medical doctors collaborated \cite{Giordano}, and the work by Fokas, Cuevas-Maraver and Kevrekidis \cite{Fokas}, in which such a model -- involving five compartments like the present paper, but chosen in a different way -- is used to discuss (as in the present paper) exit strategies from the COVID lockdown.

As mentioned in the main text, one could -- and should -- consider epidemic dynamics \emph{on networks} \cite{Newman}. Attempts to analyze the COVID epidemic in this way have of course been pursued, both on a small scale, with a network structure which can be determined by direct sociological study \cite{Anita}, and on a nationwide scale \cite{Gatto} where the network structure has to be determined.

This latter study \cite{Gatto} also attempted to evaluate the effect of the containment measures; such a matter is of course very relevant and has been considered by many authors in many countries; even a cursory mention of these is impossible, and we will just mention one study applying to Italy \cite{Guzzetta}. We also stress that many of the papers mentioned above, see in particular \cite{Giordano,Fokas} aim at using the models they study to evaluate the effect of interventions and containment measures.

Finally we would like to end on a positive note, and mention that while on the one hand it was found that the presence of asymptomatic makes that the basic reproduction number of COVID is higher than initially estimated \cite{GBRN,Gasir,DSP}, the fact that the social contact rate is not uniform in the population makes that the herd immunity level should be lower than predicted on the basis of the standard SIR-type models \cite{TBritton}; this is a specially nice result of the analysis on networks,as it only depends on general -- and very reasonable --properties of the network and not on its detailed structure, thus overcoming the low statistics problem mentioned in Remark A1 above.

\end{appendix}

\end{document}